\documentclass{article}

\usepackage{amsfonts}
\usepackage[colorlinks=true,citecolor=blue,urlcolor=cyan,linkcolor=red]{hyperref}
\usepackage{amssymb}
\usepackage{amsmath,amssymb,amscd,epsfig,amsfonts,rotating}
\usepackage{graphicx}
\usepackage{epsfig}
\usepackage{multirow}
\usepackage{booktabs}
\usepackage{url}
\usepackage[margin=1.5in]{geometry}

\newtheorem{def:def}{Definition}
\newtheorem{thm:thm}{Theorem}
\newtheorem{thm:lm}{Lemma}

\newcommand{\ctr}{r} 

\title{Optimal Real-Time Bidding Frameworks Discussion}
\author{Weinan Zhang$^{\dag}$, Kan Ren$^{\ddag}$, Jun Wang$^{\dag}$\\
{\small $^{\dag}$University College London, $^{\ddag}$Shanghai Jiao Tong University}\\
{\small \textsf{w.zhang@cs.ucl.ac.uk, kren@apex.sjtu.edu.cn}}}
\begin{document}
\maketitle

\begin{abstract}
This note is a complementary material for the solution of optimal real-time bidding function in paper \cite{zhang2014optimal}, where the estimated cost is taken as the bid price, i.e., the upper bound of the true cost. Here we discuss a more general bid optimisation framework with various utility and cost functions.
\end{abstract}

\section{Preliminaries}
Define the bidding function as $b(\ctr)$ which returns the bid price given the estimated click-through rate (CTR) $\ctr$ of an ad impression.

\vspace{5pt} \noindent \textbf{Winning probability.} Given the market price distribution $p_z(z)$ and the bid price $b$, the probability of winning the auction is
\begin{align}
w(b) = \int_0^{b} p_z(z) dz. \label{eq:win}
\end{align}

If the bid wins the auction, the ad is displayed and the utility and cost of this ad impression are then observed.

\vspace{5pt} \noindent \textbf{Utility.} The utility function given the CTR is denoted as $u(\ctr)$. The specific form of $u(\ctr)$ depends on the campaign KPI. For example, if the KPI is click number, then
\begin{align}
u_\text{clk}(\ctr) = \ctr. \label{eq:utility-clk}
\end{align}

If the KPI is campaign’s net profit, with the advertiser's true value on each click is $v$, then
\begin{align}
u_\text{rev}(\ctr) = v\ctr - z. \label{eq:utility-rev}
\end{align}

\vspace{5pt} \noindent \textbf{Cost.} The expected cost if win given a bid $b$ is denoted as $c(b)$. In RTB ad market we have first price auctions
\begin{align}
c_1(b) = b, \label{eq:cost-1}
\end{align}
and second price auctions
\begin{align}
c_2(b) = \frac{\int_0^{b}z p_z(z) dz}{\int_0^{b} p_z(z) dz}. \label{eq:cost-2}
\end{align}
In \cite{zhang2014optimal}, we used $c_1(b)$ to model the upper bound of the second price auctions with possible soft floor prices. Here we shall first use the abstract cost function $c(b)$ in the framework and then specify the implementation of the cost function in specific tasks.

\section{Bid Optimisation without Budget Constraint}

For non-budget bid optimisation, only $u_\text{rev}(\ctr)$ utility function is meaningful\footnote{If there is no cost-related factors in the utility, one will bid as high as possible to win every impression as there is no budget constraint.}
\begin{align}
U_\text{rev}(b(\cdot)) &= T  \int_{\ctr} \int_{z=0}^{b(\ctr)} (v\ctr - z) p_z(z) dz \cdot p_r(r)  dr.
\end{align}

Take the gradient of net profit $U_\text{rev}(b(\cdot))$ w.r.t. the bidding function  $b(\ctr)$ and set it to 0,
\begin{align}
\frac{\partial U_\text{rev}(b(\cdot))}{\partial b(\cdot)} = (v\ctr - b(\ctr)) \cdot p_z(b(\ctr)) \cdot p_r(r) = 0,
\end{align}
which derives
\begin{align}
b(\ctr) = v \ctr \label{eq:opt-bid-non-budget}
\end{align}
i.e., the optimal bid price is set as the impression value. Thus the truth-telling bidding is the optimal strategy when there is no budget considered.

\section{Bid Optimisation with Budget Constraint}
Assuming we want to find the optimal bidding function $b()$ to maximise the campaign-level KPI $\ctr$ across its auctions over the lifetime with total bid requests volume $T$ and budget $B$.
\begin{align}
\max_{b()} ~~ &T \int_r u(\ctr)  w(b(\ctr)) p_\ctr(\ctr) d\ctr \label{eq:obj} \\
\text{subject to}~~ & T \int_\ctr c(b(\ctr)) w(b(\ctr)) p_\ctr(\ctr) d\ctr = B \nonumber
\end{align}

The Lagrangian of the optimisation problem Eq.~(\ref{eq:obj}) is
\begin{align}
\mathcal{L}(b(\ctr),\lambda) = & \int_{\ctr} u(\ctr) w(b(\ctr))p_\ctr(\ctr)d\ctr - \lambda \int_{\ctr} c(b(\ctr)) w(b(\ctr))p_\ctr(\ctr)d\ctr + \frac{\lambda B}{T}, \label{eq:lagrangian}
\end{align}
where $\lambda$ is the Lagrangian multiplier. Based on calculus of variations, the Euler-Lagrange condition of $b(\ctr)$ is
\begin{align}
\frac{\mathcal{L}(b(\ctr),\lambda)}{\partial b(\ctr)} = 0,
\end{align}
which is
\begin{align}
u(\ctr) p_\ctr(\ctr)\frac{\partial w(b(\ctr))}{\partial b(\ctr)} - \lambda p_\ctr&(\ctr) \Big[ \frac{\partial c(b(\ctr))}{\partial b(\ctr)} w(b(\ctr)) + c(b(\ctr))\frac{\partial w(b(\ctr))}{\partial b(\ctr)} \Big]  = 0\\
\Rightarrow ~~~ \lambda \frac{\partial c(b(\ctr))}{\partial b(\ctr)}w(b(\ctr)) = &\Big[ u(\ctr) -  \lambda c(b(\ctr))\Big] \frac{\partial w(b(\ctr))}{\partial b(\ctr)}.\label{eq:general-condition}
\end{align}

Eq.~(\ref{eq:general-condition}) is a general condition of the optimal bidding function, where the specific implementations of winning function $w(b)$, utility function $u(\ctr)$ and cost function $c(b)$ are needed to derive the corresponding form of optimal bidding function.

\subsection{First-Price Auction}
With cost function $c_1(b)$ in Eq.~(\ref{eq:cost-1}), Eq.~(\ref{eq:general-condition}) is written as
\begin{align}
\lambda \int_0^{b(\ctr)} p_z(z) dz =(u(\ctr) - \lambda b(\ctr)) \cdot   p_z(b(\ctr)) \label{eq:general-condition-first-auction}
\end{align}
which still depends on the detailed form of market price distribution $p_z(z)$ to solve the $b()$. In \cite{zhang2014optimal}, the authors tried an implementation:
\begin{align}
p_z(z) &= \frac{l}{(l+z)^2},\\
\Rightarrow ~~ w(b) &= \frac{b}{b+l} \label{eq:win-1}.
\end{align}

Taking Eq.~(\ref{eq:win-1}) and click utility Eq.~(\ref{eq:utility-clk}) into Eq.~(\ref{eq:general-condition-first-auction}) we have
\begin{align}
\lambda \frac{b(\ctr)}{b(\ctr) + l} &=(u(\ctr) - \lambda b(\ctr)) \cdot   \frac{b(\ctr)}{(b(\ctr) + l)^2} \\
\Rightarrow ~~ b(\ctr) &= \sqrt{\frac{u(\ctr) l}{\lambda} + l^2} - l, \label{eq:bid-func-1}
\end{align}
which is the result in \cite{zhang2014optimal}.

\subsection{Second-Price Auction}
Taking the definition of winning function Eq.~(\ref{eq:win}) and the second-price cost function Eq.~(\ref{eq:cost-2}) into Eq.~(\ref{eq:general-condition}), we have
\begin{align}
& \lambda \frac{b(\ctr) p_z(b(\ctr)) w(b(\ctr)) - p_z(b(\ctr)) \int_0^b z p_z(z)dz} {w(b(\ctr))^2} w(b(\ctr)) = (u(\ctr) - \lambda c(b(\ctr))) p_z(b(\ctr)) \\
\Rightarrow ~~~& \lambda (b(\ctr) - c(b(\ctr))) = u(\ctr) - \lambda c(b(\ctr))\\
\Rightarrow ~~~& b(\ctr) = \frac{u(\ctr)}{\lambda}, \label{eq:opt-bid}
\end{align}
where we can see the optimal bidding function is linear w.r.t. CTR $\ctr$.

The solution of $\lambda$ is obtained by taking Eq.~(\ref{eq:opt-bid}) into the constraint
\begin{align}
T \int_\ctr c(b(\ctr)) w(b(\ctr)) p_\ctr(\ctr) d\ctr = B,
\end{align}
which is rewritten as
\begin{align}
\int_\ctr c\Big(\frac{u(\ctr)}{\lambda}\Big) w\Big(\frac{u(\ctr)}{\lambda}\Big) p_\ctr(\ctr) d\ctr = \frac{B}{T}.\label{eq:budget}
\end{align}
We can see that essentially the solution of $\lambda$ makes the equation between the expected cost and the budget. Furthermore, as both $w(\ctr/\lambda)$ and $c(\ctr/\lambda)$ monotonously decrease as $\lambda$ increases, it is quite easy to find a numeric solution of $\lambda$.

\section{Discussions}

The implementation of the cost constraint in Eq.~(\ref{eq:obj}) needs careful modelling based on the data. In \cite{zhang2014optimal} the authors used the cost upper bound, i.e., the bid price, to control the cost and let the total value of cost upper bound be the campaign budget. Here if we implement the expected cost in second price auction Eq.~(\ref{eq:cost-2}), the cost constraint in Eq.~(\ref{eq:obj}) might not be controlled by the budget. With about 50\% probability, the budget will be exhausted in advance, which is not expected in practice.

We can easily find in the first-price bidding function Eq.~(\ref{eq:bid-func-1}), the bid price is jointly determined by utility function $u(r)$, $\lambda$ and market parameter $l$.
Specifically, the bid price is monotonic increasing w.r.t. utility while decreasing w.r.t. $\lambda$.
Moreover, different value settings for parameter $l$ also influence the final bid decision as is shown in \cite{zhang2014optimal}.
As is defined in Eq.~(\ref{eq:win-1}), we can tune the parameter $l$ to alter the winning probability so as to fit different market environments.
In fact, we may conclude that: the market consideration influences bid price by tuned $l$, the budget constraint controls bid function by $\lambda$, while advertiser's utility expectation $u(r)$ could finally determine the final decision.

Let's take a look at the bid strategy under second-price auctions.
In Eq.~(\ref{eq:opt-bid}), the bid price is mainly determined by utility $u(r)$.
However, the bid strategy constructs a bridge between utility and budget consideration by $\lambda$.
And let us move our attention onto Eq.~(\ref{eq:budget}) and we can find that $\lambda$ is also controlled by this equation which takes both estimated CTR $r$ and winning probability $w(\cdot)$ into consideration.
But the latter two factors have low effects on bid strategy through $\lambda$.

From the discussion above, we may find that both bid strategies under either auction setting are influenced by three factors: advertiser's expected utility, budget constraints and market information.
While under first-price auction setting, the bid strategy takes utility function and market price together, but the strategy upon second-price auctions reflects utility more straightly.

For the bidding function under first-price auctions, we could take more attention on the winning probability estimation problem and consequently to optimize the bid strategy.
We also think that it could be more crucial to take the market information into consideration in the utility function for second-price auctions.
Finally, we may take a step forward that, to coordinates the optimization for both CTR estimation and bidding strategy, considering market information and budget capping, to dynamically bid in real-time and real-world environments.

\bibliographystyle{abbrv}
\bibliography{bid-opt-frame}

\end{document}